# A Customized Memory-aware Architecture for Biological Sequence Alignment

NASRIN AKBARI, University of Tehran
MEHDI MODARRESSI, University of Tehran
ALIREZA KHADEM, University of Tehran

Sequence alignment is a fundamental process in computational biology which identifies regions of similarity in biological sequences. With the exponential growth in the volume of data in bioinformatics databases, the time, processing power, and memory bandwidth for comparing a query sequence with the available databases grows proportionally. The sequence alignment algorithms often involve simple arithmetic operations and feature high degrees of inherent fine-grained and coarse-grained parallelism. These features can be potentially exploited by a massive parallel processor, such as a GPU, to increase throughput. In this paper, we show that the excessive memory bandwidth demand of the sequence alignment algorithms prevents exploiting the maximum achievable throughput on conventional parallel machines. We then propose a memory-aware architecture to reduce the bandwidth demand of the sequence alignment algorithms, effectively pushing the memory wall to extract higher throughput. The design is integrated at the logic layer of an emerging 3D DRAM as a processing-in-memory architecture to further increase the available bandwidth. The experimental results show that the proposed architecture results in up to 2.4x speedup over a GPU-based design. Moreover, by moving the computation closer to the memory, power consumption is reduced by 37%, on average.

## 1 INTRODUCTION

Pairwise sequence alignment is the most widely used operation in the computational biology [1][2][3]. With the exponential growth of the size of the genomic data and DNA sequence databases, which has been doubling every 7 months since 2001[2], bio-data analysis would require excessive processing power and memory bandwidth.

This exponential growth in bio-data size has been primarily enabled by the modern high-throughput sequencing technologies. These technologies allow concurrently producing thousands to millions of long reads of contiguous DNA sequences, required for reference-guided or unreferenced assembly of the human (and other organisms) genome.

Sequence alignment is the most time consuming step of these sequence assembly methods, which assemble short nucleotide sequences into longer ones to determine the precise order of nucleotides within a DNA molecule. In addition to genome assembly, pairwise global sequence alignment is also the most basic operation in biological database homology search that aims to find the most similar DNA of a database (that can be at the order of hundreds of millions of nucleotides long) to a given query sequence [1][3].

Among various solutions for the sequence alignment problem, dynamic programming (DP) methods have gained more popularity [1][3][4][5], mainly because they make a better compromise between accuracy and speed.

DP can be either applied directly on the entire length of the two sequences or on some selected parts of the sequences specified by a pre-processing step [3]. While full-length search guarantees finding the optimal alignment, heuristic methods trade optimality for faster execution. In either case, DP-based implementations consist of simple and low-precision integer arithmetic and logic operations. Consequently, DP methods feature low operational density (operations per byte), in that the amount and complexity of the operations that run on each block of data is low. However, these algorithms work on a huge amount of data: biological sequences can be very long and the size of DNA sequences can range from a few to hundreds of millions of nucleotides (characters) [1][6]. Further, the genomic databases used for homology search are growing in size, with major widely-used public databases currently containing up to 245 billion characters [2][10].

This low operational density, coupled with poor data cache utilization of the DP algorithms, makes the DP-based sequence alignment methods memory-bound; in that most of the execution time is spent reading and writing data.

The simple logic and intrinsic parallelism of the DP-based sequence alignment makes them amenable to acceleration by highly-parallel machines. Prior work have accelerated alignment algorithms on various parallel hardware platforms, including ASIC or FPGA-based application-specific architectures, graphical processing units (GPUs), and high-performance clusters [1][5][6].

However, the bandwidth demand of the algorithm increases proportional to the degree of parallelism and practically limits the throughput of the sequence alignment accelerators. In other words, although we can employ thousands of simple cores to run the algorithm in parallel, it is the limited insufficient memory bandwidth, rather than the processing power, that confines the maximum achievable throughput [7-9].

In our previous work, we showed that the throughput of the sequence alignment algorithms is highly sensitive to the bandwidth [9]. In this paper, we present a memory-aware architecture for the sequence alignment problem to push this bandwidth wall, effectively extracting much higher potential parallelism (and hence throughput) from the DP-based alignment algorithms.

To reduce the stress on memory bandwidth, the proposed architecture (1) applies a stream-based processing scheme to reuse partial results and reduce memory accesses and (2) is customized for in-memory implementation as a processing-in-memory (PIM) unit.

The PIM-based design follows the emerging 3D memory architectures in which a logic layer (consisting of memory controller units and customized processing units) is stacked on top of multiple layers of DRAM in a 3D fashion. Some major memory chip vendors already ship emerging 3D-memories with integrated logic layer [11]. The logic layer implements a memory controller and often some simple functional units to execute a set of in-memory instructions [11].

In such architectures, the memory-intensive code segments are offloaded to the logic layer of the 3D memory to minimize data movement between the processor and memory chips. The offloaded code takes full advantage of the low-latency and high-throughput data delivery of the short and fast 3D vertical links.

To approach the ideal throughput of the sequence alignment algorithm, the contributions of this paper are twofold. We fist show that the throughput of the DP algorithms on conventional processing units, i.e. multicore CPUs and GPUs, is highly bound by memory bandwidth: when the available bandwidth grows, throughput will scale accordingly. Second, we design a memory-aware PIM accelerator architecture for the DP-based sequence alignment algorithms to bridge the gap between the GPU/CPU's and ideal throughputs.

The experimental results show that the memory-side design considerably increases the performance and reduces the power consumption when compared to conventional high-performance sequence alignment methods and architectures.

The rest of the paper is organized as follows. Section 2 covers preliminary background. Section 3 reviews some recent related studies. Section 4 presents the motivations behind designing a memory-side accelerator for sequence alignment through a quantitative study. Section 5 describes the proposed architecture. Section 6 presents the proposed PIM-based design. Section 7 outlines the implementation and evaluation methodology, followed by experimental results in Section 8. Finally, Section 9 concludes the paper.

## 2 BACKGROUND

A Customized Memory-aware Architecture for Biological Sequence Alignment

*a. Pairwise sequence alignment*

A biological sequence is a chain of nucleotides in a DNA or amino acid residues in a protein [5]. This paper targets DNA sequences, in which very long sequences are made of only four types of nucleotides. We refer to *nucleotides* as *characters*, hereinafter.

Pairwise global sequence alignment is an important and challenging task in bioinformatics that is used to identify regions of similarity between a given query sequence and a set of database sequences. This similarity is used to explore structural, functional and evolutionary relationship between the sequences [1][3]. The well-known DP approach divides the problem into smaller independent sub-problems and explores the possible alignments in a quantitative way. It compares the sequences element-wise and gives some certain scores for matches and mismatches of each pair of characters, which are then summed up to get the alignment score of the two sequences.

The Needleman-Wunsch algorithm is perhaps the most widely-used dynamic programming solution for the pairwise sequence alignment problem. It compares two sequences by computing a distance value representing the minimal cost to transfer one subsequence into another [3].

Let us consider two sequences $A=(a_1 a_2 \ldots a_m)$ of length $m$ and $B=(b_1 b_2 \ldots b_n)$ of length $n$. The algorithm computes an optimal alignment between two sequences by filling a dynamic programming matrix $DP(i,j)$ of size $(m+1) \times (n+1)$, from cell $(0,0)$ to cell $(m,n)$ in a row-by-row manner (see Figure 1.a).

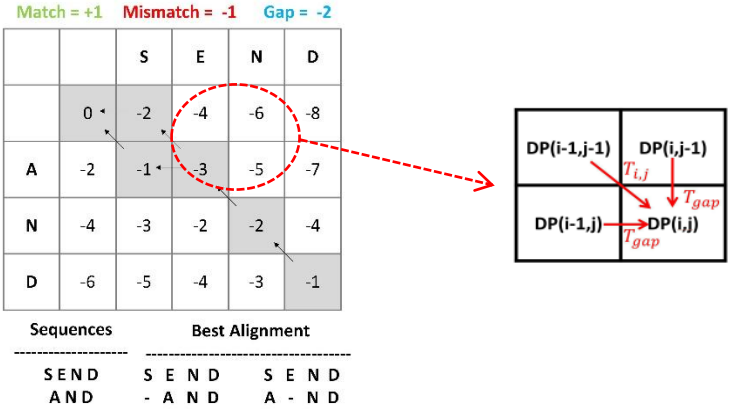

Fig. 1. (a) Dynamic programming matrix for alignment of "SEND" and "AND" sequences (the arrows show backward pointers) (b) Maximum score propagation flow

In the DP matrix, the value of each cell is computed based on northwest diagonal, north, and west immediate neighboring cells (see Figure 1.b). Three separate scores are calculated from these neighbors and the maximum score is assigned to the cell as:

$$DP(i,j) = MAX \begin{cases} DP(i-1, j-1) + T_{i,j} \\ DP(i-1, j) + T_{gap} \\ DP(i, j-1) + T_{gap} \end{cases} \quad (1)$$

Where $T_{i,j}$ is the similarity score of comparing $a_i$ to $b_j$ and $T_{gap}$ is the penalty for inserting gap in either sequences.

A sequence may be derived from a parent sequence by insertion/deletion of some characters. To take this phenomenon in the alignment procedure into account, the strings can be extended by so called gap (blank) characters, which can be inserted at any position in the strings to get a better alignment. The sequence alignment problem then aims at finding the best alignment (with the highest possible score) of two sequences of characters by appropriately inserting the gap characters in either sequence.

Figure 1 shows the alignment of two sequences SEND and AND, with the gap penalty of -2, mismatch penalty of -1, and match score of +1.

b. *Processing in memory*

We model our processing-in-memory architecture after the Micron's Hybrid Memory Cube (HMC), which is a commercial implementation of a 3D stacked DRAM with one integrated logic layer [12][14]. This 3D design integrates one logic and multiple memory dies on top of each other in a 3D fashion. The logic layer benefits from the abundant memory bandwidth provided by the high-speed vertical inter-layer links made by the Through-Silicon Via (TSV) technology. HMC has up to 8 DRAM layers for a total of 8GB capacity. Each DRAM layer is divided into 32 partitions and all vertically adjacent partitions, which form a column of stacked partitions, are called a *vault*. Each vault has an independent DRAM controller, or vault controller, implemented on its logic die. Having private memory controllers, vaults act as independent memory channels and can be accessed simultaneously.

HMC is connected to the host processor through up to four 16-bit full-duplex serial links on which, the memory requests/responses are transferred in the form of packets. The logic layer at the memory chip provides a 4×32 crossbar switch that directs packets to the right vault controller.

In addition to implementing memory controller, the HMC 2.0 standard implements several simple in-memory arithmetic and logic units in its logic layer. These units are invoked by corresponding instructions at the host processor. In this paper, we extend the PIM capability of this architecture by adding our proposed sequence alignment accelerator unit to it.

While implementing a full-fledged processor in the logic layer of a 3D memory is not currently practical due to some serious thermal issues, prior work on PIM show that using power-efficient accelerators for this purpose is a viable approach that will not cause severe thermal issues [13][14][16]. This way, most existing PIM proposals use memory-side accelerators tailored to a specific application/domain. These accelerators are controlled by the host processor to perform specific bandwidth-intensive tasks in memory.

There are two main reasons behind the power-efficiency of accelerators. First, accelerators just implement the target datapath, without any extra logic that a general-purpose processor would have to provide for the sake of flexibility. Second, accelerators eliminate the software-related pipeline stages (fetch, decode, and scheduling). In some previous work, it has been shown that accelerators can save up to 90% of the power usage of the computation [17].

By moving accelerators to the memory-side, we further reduce the power-hungry on-board processor-to-memory data migration and hence, can reduce the total system power consumption more.

## 3 RELATED WORKS

There are several hardware accelerators for sequence alignment in the literature [1].
FPGAs are one of the most widely-used platforms for accelerating the sequence alignment algorithms [18][20][22][23]. For example in [18], it has been shown that the processing time of the Smith-Waterman alignment algorithm on FPGA can be improved by an average of 287%, compared to the pure software implementation. However, the FPGAs suffer from the memory bandwidth limitation, which stems from the limited on-chip storage capacity. GateKeeper is another example that accelerates the alignment problem on FPGAs [23]. In this method, the performance of genome read mapping has been increased substantially by filtering out incorrect candidate locations in the reference sequences.

In [22], the authors have proposed a hardware-acceleration framework for genome sequence alignment, named *Darwin*, which accelerates the compute-intensive tasks of the long-read assembly procedure by orders of magnitude. Implemented on FGPAs, Darwin aims at aligning long reads to a reference genome without sacrificing sensitivity. It uses a novel filtration heuristic and a novel alignment algorithm based on dynamic programming. Our proposed PIM design is a general method that provides more memory bandwidth for the alignment problem, so can be integrated with Darwin to further increase performance.

In several prior work, a customized network-on-chip (NoC) is designed to carry the inter-core traffic in many-core implantation of the Needleman-Wunsch sequence alignment algorithm [24][25]. Focusing on the inter-core communication, they do not address the off-chip memory limitations.

In addition to custom hardware designs, a large body of research and practical works accelerate the sequence alignment on GPUs [19][21][26] and high-performance processors [27][28].

In [26], a parallel implementation of the Needleman-Wunch algorithm on GPU is presented. The evaluation results shows up to 99x speedup over a conventional processor. The speed of the Smith-Waterman algorithm on Intel Xeon Phi is also shown to be 10-100x faster than a conventional desktop processor [27].

Other hardware architectures that are found useful to speed up the alignment procedure include IBM's Cell BE [29] and the SIMD units of traditional CPUs [30].

In a previous work, however, we showed the performance of these massive parallel CPU and GPUs is bound by the memory bandwidth and by moving the processing closer to the memory more throughput can be archived [9]. In this paper, we extend our early study by proposing a memory-aware PIM architecture to push the memory wall towards exploiting the maximum algorithm parallelism.

Recently, some papers adopt the emerging resistive RAM (ReRAM) technology to implement alignment algorithms [2][31]. These methods exploit the interesting capability of ReRAM in building area-efficient associative memory. In general, data replication is the drawback of these ReRAM-based mechanisms. As a future work, we can consider implementing the proposed memory-aware accelerator on ReRAM.

## 4 SEQUENCE ALIGNMENT AND THE MEMORY WALL

According to Equation 1, each cell of the DP matrix is calculated by a single comparison of 2-bit sequence characters followed by a single 32-bit integer addition operation. This involves fetching two 2-bit characters and two 32-bit DP cells and writing back the 32-bit result (which is an updated DP cell) to the memory. A simple rule of thumb suggests that filling the DP matrix serially (with a rate of one cell per cycle) in a processing element with 1GHz working frequency requires 100Gb/s (12.5 GB/s) memory bandwidth. By parallelizing the alignment process, the bandwidth demand grows proportional to the parallelism degree. Current parallel methods exploit either coarse-grained parallelism (by concurrent alignment of a query sequence to multiple sequences of a database in parallel) or fine-grained parallelism (by parallelizing the alignment algorithm itself) to accelerate the sequence alignment process [1].

Generally, data cache can filter part of the accesses. However, the Needleman-Wunsch algorithm, like most bioinformatics algorithms, exhibit poor temporal locality, because once a reference sequence is fetched, it will never be used again for the current query.

DP matrix cells also suffer from a large reuse distance that is proportional to the sequence length. Once a cell is calculated and written to the DP matrix, it will be read again to calculate the cells of the succeeding row, but this will happen when the algorithm runs over the entire DP row (which may have thousands of cells). For large sequences, it is very likely for the cache to replace the cell's data before the actual reuse.

Abundant memory bandwidth demand, coupled with poor caching behaviour, lengthens the alignment process, particularly when a query is to be aligned with a large database of reference sequences. This highlights the need for memory-aware implementation of the sequence alignment algorithms.

**Impact of memory bandwidth.** As a quantitative motivation for the need for such memory-aware mechanisms, we show how general-purpose GPUs fall short of exploiting maximum parallelism due to bandwidth limitations.

A modern GPU assembles hundreds to thousands of processing cores capable to execute thousands of parallel threads. GPUs, as mentioned before, have been used to solve the sequence alignment problem in several previous studies [1].

To evaluate the impact of bandwidth on the performance of GPUs when running the sequence alignment algorithm, we compare the execution speed of a GPU with realistic and ideal memory bandwidths.

We use GPGPUSim [32] with the Nvidia's Fermi architecture to align 60k-character sequences. We use a publically-available implementation of the Needleman-Wunsch algorithm with CUDA [33]. The GPU is configured with 15 Stream Multiprocessors (SMs) and 32 Stream Processors (SPs) per SM for a total of 480 SP cores. The cores work at 1.4 GHz.

The Fermi architecture uses 64KB L1 private cache per SM and 768KB L2 shared cache to store recent calculated values. The architecture supports 6GB GDDR5 DRAM with the bandwidth of 177GB/s.

To investigate the effect of bandwidth on the achievable performance, Figure 2 compares the throughput of the GPU with the realistic 177GB/s and the ideal infinite bandwidths. The comparison metric is *Giga-Cell Update Per Second* (GCUPS). GCUPS is a common performance measure used in computational biology that shows how many DP matrix cells are updated (calculated) per time unit.

The results, depicted in Figure 2, indicate a considerable gap (around 3X) between the performance of a GPU with the limited bandwidth and the ideal. It shows that the performance is highly sensitive to the off-chip bandwidth and expected to grow with the available bandwidth proportionally.

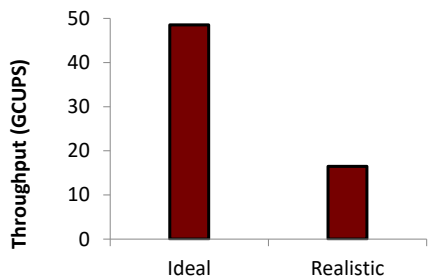

Figure 2. Impact of bandwidth on the throughput of a GPU

Current CUDA code just exploits the internal parallelism of the algorithm in a fine-grained manner. Obviously, by increasing the degree of parallelism, by aligning a query sequence with multiple reference sequences in parallel, more bandwidth is needed.

Figure 3 confirms our previous analytical statement that the long reuse distance of the algorithm cannot be captured by reasonably sized L1caches.

A Customized Memory-aware Architecture for Biological Sequence Alignment

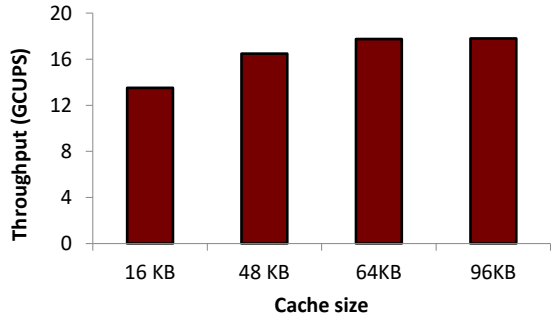

Figure 3. Impact of cache size on the throughput of a GPU

**Impact of cache size**. To investigate the effect of cache on the achievable speedup, we repeat the simulations with four valid L1 cache sizes of the considered GPU architecture. The size of L2 cache is 768KB and memory bandwidth is 177GB/s.

**Impact of memory management methods**. In addition to large caches, prefetching and sequence compression are two more ways to increase the performance of the memory system. Sequence compression techniques, which try to reduce sequence size, are orthogonal to our method and can be applied along with it to further increase performance.

Prefetching is a prediction-based mechanism that accelerates memory access by bringing data to cache before the actual access. However, prefetching is a latency reduction mechanism: it reschedules memory accesses to reduce latency, but since it do not change the bandwidth demand, cannot push the bandwidth constraints. Therefore, it cannot increase the performance of bioinformatics algorithms, where we mainly care about the throughput (alignment per second) rather than the speed of a single alignment operation.

In this paper, we design a memory-aware architecture for the Needleman-Wunsach alignment algorithm that reduces the bandwidth demand by reusing partial results. To further push the bandwidth wall and narrow the gap to the maximum achievable performance, we move the alignment computation closer to the memory by integrating it to the logic layer of a 3D memory.

## 5  THE PROPOSED ACCELERATOR ARCHITECTURE

Figure 4 shows the dataflow and datapath of the accelerator designed for the sequence alignment problem. While a naïve implementation would fill the matrix row-by-row and one cell at a time, our design uses a property of the DP matrix that allows filling multiple cells diagonally at the same time.

Figure 4.a shows this scheme. At each step, all cells at the boundary of the filled cells (whose values are updated) and unfilled cells can be updated in parallel in a wavefront manner, as they have all the required input cells already calculated. These cells are diagonally adjacent, with the number of diagonal cells vary from 1 (when calculating the first and last cells of the DP matrix) to the size of the diameter of the DP matrix.

Obviously, providing sufficient functional units to exploit the maximum parallelism is not feasible, due to the large area requirement. It also leads to frequent resource underutilization, as the parallelism degree is less than the maximum most of the time.

For a reference sequence of size $n$ (Sequence A in Figure 4.a) and query sequence of size $m$ (Sequence B in Figure 4.a), if we have $p$ functional units to calculate $p$ cells simultaneously, our accelerator partitions an $n \times m$ DP matrix horizontally into $n/p$ blocks of size $p \times m$. Each block is filled in $m$ cycles (starting from the leftmost column) and at each cycle, the part of the diagonal cells that lie within the block are filled in parallel, as illustrated in Figure 4.a.

The blocks are processed in serial. Starting from the uppermost block of the DP matrix, filling the entire DP matrix will be completed in $n/p$ steps (each involving $m$ cycles to pass over the entier row). We can process the blocks in parallel, but the parallel processing involves exchanging some intermediated data. By exploiting coarse-grained parallelism, i.e. by comparing a query sequence to multiple references, we can keep any number of processing elements busy without incurring the communication overhead of the parallel algorithm.

Figure 4.a shows this block-based cell updating flow with the block width set to four. Particularly, the figure highlights three diagonal cells to shows how the calculation of each set of diagonal cells in each block (the rightmost highlighted diagonal in Figure 4.a) depends on the values of the two most recent calculated diagonals.

To calculate the value of a single cell, the corresponding functional unit should be provided by the value of the three surrounding north, north-west, and west cells.

As Figure 4.a indicates, during the calculation of a diagonal, the values of the surrounding cells come from the two recent calculated diagonals. The only exception is the uppermost cell of each block (located in the first row of the block) which should fetch the value of its north and north-west cells from memory. These values come from the lower row of the most recently processed block. Of these two values, the north-west cell has already been fetched as the north cell of the preceding cell of the row (whose value is updated in the previous cycle), so already exists in the accelerator. Figure 4.c displays how the north neighbor of cell $X$ is reused as the north-west cell when calculating the value of the next cell (cell $Y$). Thus, the accelerator just needs to fetch a single value for the uppermost cell of a diagonal.

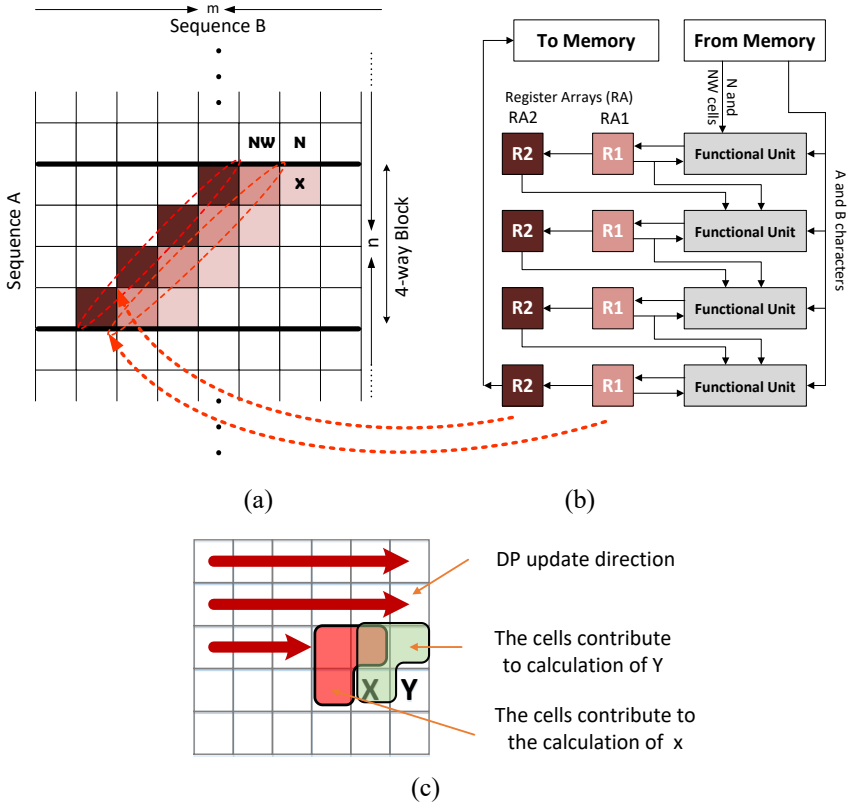

Figure 4. (a) The block-based updating flow of DP, (b) the block-based architecture, and (c) different roles of a cell to calculate two consecutive DP cells X and Y

A Customized Memory-aware Architecture for Biological Sequence Alignment

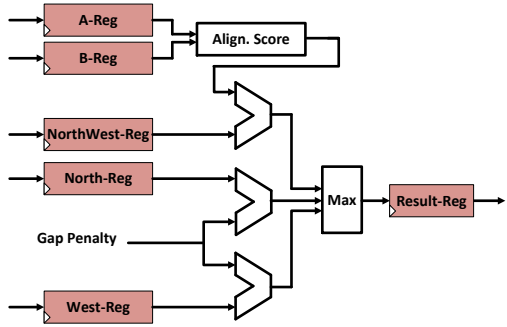

Figure 5. The structure of the functional unit

The accelerator also needs the respective characters of the reference and query sequences. The functional units are initialized with 4 characters of sequence *A* at each step, which are read once for each block and remain stationary in the accelerator as long as the current block is under process. The characters of *B*, however, are streamed to the functional units one per cycle.

Figure 4.b shows the architecture of the accelerator with four functional units. It consists of one array of four functional units and two arrays of four registers. At any cycle *t*, the register arrays *RA1* and *RA2* hold the values that are recently calculated in cycle *t-1* and cycle *t-2*, respectively. At each cycle, the contents of *RA2* are written back to the memory, the contents of RA1 are shifted to *RA2*, and *RA1* is updated with the values of the currently processed diagonal.

Please note that we only need to write back the lowermost *RA2* register to memory, since it is needed to calculate the value of some cells of the next block. The values of the other *RA2* registers have been used internally and will not participate in any calculations in the subsequent blocks.

Figure 5 shows the internal structure of the functional units. According to Equation 1, this functional unit can be implemented by a few comparators and adders.

As Figure 5 indicates, at each cycle, two characters from the two sequences under comparison and the three required DP cell values are received from the five input ports of the functional unit. The two ports *A* and *B* carry the string characters and the required DP cells come through ports north, north-west and west.

In Figure 5, the "alignment score" block calculates the matching score of each pair of characters. At its simplest form, it returns 1 for a match (when two characters are identical) and -1 for mismatch and matching with a gap.

**Memory requirements**. This architecture keeps the DP cells in the PE registers, only as long as they are needed, hence do not need to keep the entire DP matrix in memory. The lowermost row is the only part of each block that should be kept in memory, in order to be used to calculate the uppermost row of the next block.

This way, to align two sequences of size *n* and *m* and block width of *p*, which divides the $n \times m$ DP matrix into *n/p* blocks of size $p \times m$, we need to: (1) write and read the *m* boundary cells of the blocks (*n/p*)-1 times (for a total of ((*n/p*)-1)×*m* 32-bit read and write operations), and (2) read the entire elements of sequences A and B with *m* and *n* characters of size 2-bit, respectively.

This amount of memory access, although by far smaller than those of the basic memory-agnostic architectures [9], may still impede exploiting the maximum throughput for sequence alignment, since it exhibits poor caching behavior. In order to boost the remaining memory accesses, we propose to implement the design as a memory-side accelerator.

# 6    IN-MEMORY ACCLERATIOR IMPLEMENTAION

**Execution flow overview**. In our design, the reference sequences of the target database are read from a storage device and are distributed across the vaults of a 3D memory at an initialization phase. The sequences are stored in consecutive addresses. A piece of metadata is associated with the database that keeps track of the number of sequences and the length of each sequence.

For each run of the database search, all PEs receive the same query sequence, by writing a separate copy of the query to some address in their vaults. Then, the host processor initiates the alignment process in all vaults concurrently to align the query sequence to the part of the database assigned to them. Each vault finds the maximum alignment score among the sequences of its database.

Afterwards, the host processor reads the local maximum score from all vaults and calculates the global maximum. Note that the PEs just implement the score calculation/propagation steps of the Needleman-Wunsch algorithm. Once the sequence with the highest score is found across all vaults, the complete algorithm, including the backward pointer calculation step, is performed on this selected sequence again to find the alignment path. This way, the backward algorithm steps required to find the alignment path is called only one time for the reference sequence with the highest score.

**In- memory implementation overview**. Figure 6 outlines the big picture of the architecture of the proposed PIM design.

In a 3D memory architecture, host communicates with the memory by exchanging packets. A crossbar in front of the memory directs packets to the right vault based on the address of the transaction encapsulated in the packet.

In our design, the accelerator described in Section 5 is implemented as a processing element (PE) on the logic layer of the vaults. Associated with each PE, the design also needs a programmable DRAM address generation unit (AGU). The PE is configured and controlled by the AGU. AGU generates a sequence of addresses to fetch the data needed by PE. It also generates required addresses to write the calculated cells back to the memory.

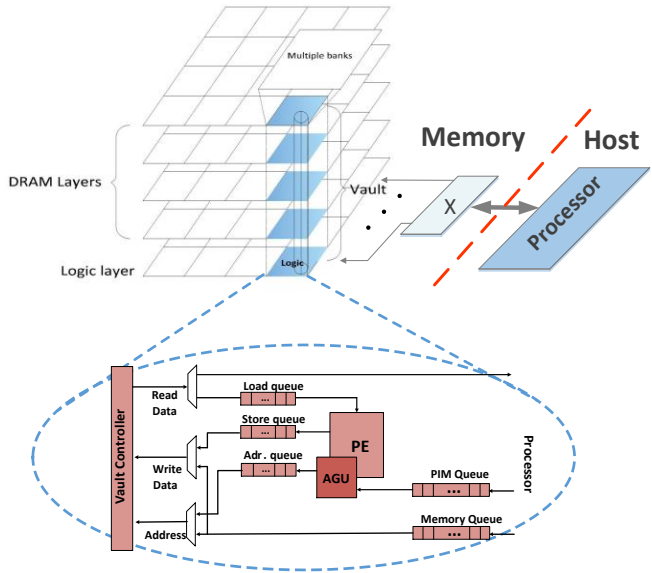

Figure 6. 3D memory and the internal structure of the logic layer of a vault

In order to generate the right sequence of addresses, AGU is programmed by the host processor through special PIM packets. PIM packets contain the AGU programming data and are tagged with

A Customized Memory-aware Architecture for Biological Sequence Alignment

the target vault number. Based on this tag, PIM packets are directed to the right vault by the input crossbar of the memory.

At each vault, the PIM packets are separated from conventional read/write packets by forwarding them to *PIM_Queue* to be processed by the AGU, whereas the regular memory requests are directed to the *Memory_Queue* in Figure 6.

By receiving each PIM packet, the AGU generates a sequence of memory access requests. These requests have the same format as the standard requests issued by the host processor for regular read/write transactions, but are placed in separate queues. An arbitration module multiplexes memory access between the regular and AGU memory access requests.

a. Address generation unit

The AGU generates the sequence of memory addresses required to calculate the value of DP cells and manages data movement between the memory and PE. The AGU, as mentioned before, is programmed by a special PIM packet received from the host.

The PIM packet should specify five parameters: the addresses of the reference and query sequences, the address of the metadata that keeps the number and the length of the database sequences, the length of the query sequence, and the starting address of the location where the DP matrix cells are stored. If $n$ reference sequences are stored in a vault, each packet initiates the pairwise alignment $n$ times, each between the query and one of the reference sequences of the vault.

At each step, AGU reads a packet from the *PIM_Queue* and generates the sequence of the required addresses according to the pseudocode displayed in Figure 7. The pseudocode is designed based on the computation flow depicted in Figure 4. This pseudo code can be easily implemented by a state machine.

For the sake of readability and in order to focus on the main concepts of the address generation algorithm, we make the following simplifications in the pseudo code. First, although the matrix-based structures are stored as a linear row-major array in the memory, we use two-dimensional addressing in the pseudo code. Second, DP algorithms add one column and one row to the DP matrix to keep track of the gap characters (see the DP matrix in Figure 1), but we ignore them in the pseudo code. Third, each memory access returns a 32-bit word [11], so we should read a new word from the DNA sequences every 16 cycles (as each DNA character can be encoded in two bits). The pseudo code makes one access per sequence character that will be reduced by a factor of 16 in real the implementation. Note that the DP matrix cells are 32 bits wide, so one separate access is required to fetch every single cell.

The DP matrix is zero-padded with $p$ columns (p=block width) in order to treat the boundary cells in the same way as the other cells. The proposed architecture just keeps two rows of the DP matrix: the lowermost row of the upper block (DP1 in the pseudocode) and the lowermost row of the current block (DP2 in the pseudocode), which is later required for the calculations of the next block. Other rows are reused internally, so are not needed to be stored in memory.

The addresses generated by the AGU are formatted as a conventional memory request and are queued in *Adr_queue* of the vault controller to be serviced in a FIFO manner (Figure 6).

For write requests, the addresses generated by AGU are associated with the contents of *Store_queue*. The *Store_queue* is filled by the PE with the corresponding DP cells values that should be written back to memory.

The data returned for each read request is written to the *Load_queue*.

As the AGU requests are prioritized over the regular read/write requests (issued by the programs at the host processor), the data will return at the same order as the requests are generated and queued.

```
main()
{
  p= block width;
  while (PIM_Q.empty())     //wait while the queue is empty
```

```
 pim_packet=PIM_Q.dequeue(); //read a packet from queue
 db_size=pim_packet.extract_no_db_seq(); //No. of database ref. sequences
 a=pim_packet.extract_address_q_seq(); //Address of query sequence
 n=pim_packet.extract_length_q_seq(); //Length of the query sequence
 dp= pim_packet.extract_dp_address(); //Address of the DP matrix
 FOR i=1 to db_size   // For all reference sequences
    m= get_next_ref_seq_length();   //get the length of the next reference sequence
    b = get_next_ref_seq_address();  //get the address of the next ref. sequence
    process_seq (a,b,n,m,p,dp);
 END FOR //i
}//main

process_ seq (a,b,n,m,p,dp);
{
 FOR (i=0 to ⌊n/p⌋) //for all blocks
    DP[0]=DP[1];
    FOR (j=1 to p)  //read the reference sequence cells
       B_Reg[j]=B[p*i+j]; //generate address to read from B and  pass the received data to Reg_B of PE
    END FOR//j
    FOR (k=1 to m)
       A_Reg=A[k];  //generate address to read from A  and pass the received data to Reg_A of PE
       North_Reg = DP[0,k];   //generate address to read data from DP to pass to North_Reg of PE
       Wait For one clock;         //wait for PE to process data
       DP[1,k-p] = Result-Reg;    //generate address to write back  to DP
    End FOR //k
  End FOR  //i

}//process_seq
```

Figure 7. AGU description

This order is exactly the same as the order at which the PE reads and writes data. As a result, the PE can correctly match the received data with datapath input ports at each cycle.

In each iteration of the PE's operation, AGU just needs to fetch the values for two registers, i.e. the North register of the first row of the block and the next character of the reference sequence to fill register B. The other required values are forwarded to the destination register internally.

Once the two requested data items are ready, AGU asserts a ready signal to pass them to the datapath and initiate data processing. Then, as the latency of datapath is one cycle, AGU generates appropriate addresses to associate it with the output of the datapath and write it back to the memory.

b. *Memory management and program control at host*

Like some prior work, the address range of sequences is marked as non-cacheable to prevent the potential cache coherency problems [14].

The programs on the host work with virtual address, but PEs work with physical addresses. To program the AGUs, the processor simply translates the virtual address of the target sequences to physical address by accessing its TLB.

Moreover, in order to write (and then access) the reference and query sequences to the right vault, we need to work with the physical address range of each vault. A prior work [15] shows that this can be done by extending the *numa_alloc_onnode()* function in Linux, which is originally developed to allocate memory on a specific node in a NUMA architecture.

The memory-side accelerators in our design are passive functional units that are controlled by the host processor. Thus, the entire algorithm runs as a single program on the host processor, but the sequence alignment algorithm is offloaded to the memory-side accelerators.
This way, unlike memory-side full-fledged processors [15][37], this design do not need complex synchronization, coherence, and consistency checks between the memory-side logic and host processor.

## 7 EXPERIMENTAL METHODOLOGY

*a. Simulation environment*

In order to evaluate the efficiency of the proposed PIM-based alignment design, we implemented an in-house extension of the Ramulator [35] memory simulator. In the simulator, the memory structure and timing parameters are set based on the HMC 2.0 specifications [11-14].

The simulator models the host at a high abstraction level and as an entity that sends PIM packets to program the AGU and initiate processing at the memory-side PEs. It runs no other program during the alignment algorithm execution, so all memory accesses are made by the alignment algorithm.

We compare the results with the GPU described in Section 4 (using GPGPUSim) and a quad-core processor that models the Xeon processor, using the MARSSX86 simulator [45]. The CPU has four out-of-order cores that work in 1.4 GHz, with 128KB private L1 and 2MB shared L2 caches. Another configuration selected for the comparison purpose is the memory-side accelerator design we proposed in [9]. It updates a single cell per cycles and does not reuse intermediate data, but relies on coarse-grained parallelism to increase throughput.

The results are also compared to the case where the PE are implemented at the processor side and receive data from an off-chip memory (with the bandwidth listed as "external bandwidth" in Table 1). We refer to this configuration as processor-side. Note that this configuration also adopts the 3D HMC-like memory, but the PE is implemented at the processor side.

As mentioned before, it is assumed that the reference sequences are read from the target database and distributed across the physical address space of the memory in advance.

The key driver for the efficiency of the memory-side acceleration is the ratio of the available internal to external memory bandwidths. The external bandwidth is the bandwidth that memory provides to the host processor via external memory-to-host links. The internal bandwidth is the memory bandwidth available to the PEs at the logic layer.

The internal bandwidth of a 3D memory depends on the number of vaults, number of TSVs per vault, and the bandwidth of each individual TSV. Some recent PIM designs in the literature consider bandwidths from 160 GB/s to 512 GB/s [14][15][34][36]. The external bandwidth can also vary significantly based on the number of links, link speed, and packetization overhead. Recent works consider different configurations with the bandwidth varying from 40 GB/s to 320 GB/s [14][15][37][38]. To study the impact of bandwidth, we evaluate the sensitivity of the speedup to different internal to external bandwidth ratios.

Table 1 outlines the considered HMC parameters. In our simulator, the internal and external bandwidths are parameterizable. We adopt the HMC.20 configuration, which provides 10 GB/s bandwidth per vault (for a total internal bandwidth of 320 GB/s per memory module) and a maximum external bandwidth of 240 GB/s (with four 16-bit 60GB/s links). The access size through external links is 32 bytes. Based on this access size and the 8-byte header and tail of packets in HMC, the packetization imposes 27% average protocol overhead on the external bandwidth. This overhead is inevitable due to the packet-based nature of 3D memories and corroborates the packetization overheads reported by Micron [12]. Note that even with this overhead, HMC provides higher bandwidth than the fastest conventional DDR memories.

Table 1. The memory parameters used in the evaluations [14][39]

| External bandwidth (up to 4 links) | Up to 240 GB/s |
|---|---|
| Internal bandwidth (peak per vault) | 10 GB/s |
| DRAM layers | 4 |
| Layer size | 1 GB |
| Number of vaults | 32 |
| Word Size | 32 bits |
| Row buffer size | 256B |

*b. PE implementation*

To obtain the circuit-level characteristics, we have implemented the RTL description of the accelerator in VHDL. The description is synthesized with a commercial synthesis tool using a publically available 15nm technology library [40].

Based on the synthesis results, the area of each row (way) of the PE, comprising one functional unit and two registers (see Figure 4.b), is 95.9um$^2$. The area of the AGU is 980um$^2$.

The area of the logic layer of HMC is 68mm$^2$ [14][43]. With 32 vaults, the area of logic layer of each vault is around 2.2mm$^2$. Even if only 10% of this area is available to our accelerator, a PE with more than 2000 ways (functional units) can be built in each vault. In this section, we use 480 functional units for fair comparison with the considered GPU, which comprises 480 SP cores. Since there are 32 vaults, the PE at each vault has 16 functional units.

According to the synthesis results, the critical path of the datapath depicted in Figure 5, is 69ps, resulting in 14.485GHz operating frequency. Under this working frequency, the average power consumption of the functional units is 0.98mW. The power consumption is extracted from the VHDL code. Since the PEs perform a monotonous task, there is no significant difference between the average and peak power consumptions.

This power consumption do not violate the maximum 300mw power budget of the logic layer of HMC [42], but may result in thermal issues, particularly in 3D architectures, where the logic and DRAM layers share the same path to the heat sink. The small area footprint of our accelerator, which increases the power density, also fuels the thermal problem.

To estimate the temperature of the memory-side PE and calculate the safe working frequency, we use the Hotspot 6.0 temperature estimation tool with its 3D stacking feature [41] under the parameters we used in a prior work [42]. The thermal simulation environment considers: (1) the under-test functional unit is surrounded by eight neighboring units of the same type (direct and orthogonal neighbors in a grid-like floorplan) in the logic layer of a 3D memory system, with the entire logic layer has a floorplan provided by Micron [43], (2) the logic layer is sandwiched between a heat sink and four DRAM layers, and (3) all functional units work at the same frequency.

The temperature is calculated based on the temperature of the neighboring functional units, the power consumption of the functional unit itself, and the temperature of the neighboring DRAM slice in the immediate upper layer. This way, the impact of elements with less significant contribution, i.e. farther functional units, TSVs, and vault controller is ignored.

As stated by the HMC 2.0 specification [12][14], the maximum safe operating temperature of the logic and DRAM dies are 383°K and 378°K, respectively.

Thermal simulation shows that the peak steady-state temperature of a functional unit when working at the highest frequency will reach to 513° K. More simulations show that we must limit the working frequency to 3.74GHz in order to keep the temperature below the 383°K threshold. So, we set the clock frequency of the memory-side PE to 3.7GHz in our experiments.

**Processor-side accelerator configuration**. We compare the results with the case where the PE are implemented at the processor side and. We refer to this configuration as processor-side. Note that this configuration also adopts the 3D HMC-like memory, but the PE is implemented at the processor side. It has 64KB cache.

The processor-side PE receives data from an off-chip memory, with the bandwidth of 240 GB/s, as listed as "external bandwidth" in Table 1. The PEs has the same numbers of functional units as in the memory-side configuration. In the simulations, the processor-side PE work at 6.6 GHz, as our temperature estimation shows that the aforementioned 383°K thermal constraint cannot be met beyond this frequency. The reason for this higher frequency is the 2D nature of the processor side implementation, which eliminates the thermal dissipation complexities of 3D designs.

*c. Workloads*

In this paper, we primarily focus on the Needleman-Wunsch global DNA alignment problem, in which a query sequence is aligned to a database of reference sequences.
In a DNA sequence, every character can take four different values, so is coded into two bits. Each DP matrix cell is 32-bit, as it should keep the cumulative alignment score, which can become quite large for long sequences as the algorithm approaches to the last cells (lower left corner of the matrix).

Since we assume that the database is already copied to the memory, the throughput results do not include the database transfer time.

Further, as the genome assembly is one of the most growing trends in computational biology, we also use the reference-guided assembly application as benchmark [22]. Reference-guided genome assembly is the computational process of reconstructing a genome from a set of so-called reads. In order to read and analyze the DNA, biological sequencing technologies break the entire genome into smaller fragments, or reads. Since humans are very similar, in reference-guided assembly, a reference DNA sequence which resembles the sequenced genome is used as the reference pattern to line the fragments (or reads) up in the correct order and reconstruct the sequenced DNA [22]. This assembly can help to find small, and with some modifications, large structural variants between the reference and the sequenced DNAs. The assembly process has the pairwise alignment between the reads and the reference sequences as the most time consuming part.

Current biological sequencing machines can extract reads with several thousands of characters. In our evaluations, we assemble 1000-character reads into a 60k-character reference genome.

## 8  EXPERIMENTAL RESULTS

### a. *Throughput evaluation*

We first evaluate the performance under the Needleman-Wunsch database search. Figure 8.a compares the throughput of the conventional processor-side and the proposed memory-side configurations with CPU and GPU. The results are normalized to the GPU to give a better insight into the obtained improvements. For reference, the throughput of the GPU is 15.9 GCUPS. The bar marked as *Memory-side conv.* is the memory-side accelerator design we proposed in [9].

As Figure 8 illustrates, the proposed PIM-based gives an average improvement of 58%, 2.4X, and 11.3X, over the processor-side accelerator, GPU, and CPU, respectively.

Note that the high-end GPUs may have more cores than the current 480 core used in our configuration, but we limit the number of functional units of the PEs to 480 for fair comparison.

The first reason of the improvement over the GPU is the stream-based customized design that removes software overhead (as all accelerators do) and considerably reduces bandwidth demand by data reuse inside the accelerator. Therefore, even the processor-side accelerator outperforms the GPU. Compared to the same accelerator architecture at the processor-side, the higher bandwidth available at the memory-side brings about higher throughput. It is evident that while memory-side accelerator incurs an overhead in terms of working frequency, this lower frequency is overcompensated by the less waiting time for data due to the larger available memory bandwidth. Having the same trend, the memory-side accelerator outperforms the competitors under the short read assembly workload (Figure 8.b). Compared to the early accelerator design we previously proposed in [9], the data reuse scheme of the current model removes the need for writing and reading the intermediate results in most cases, hence can use the saved bandwidth to extract higher parallelism.

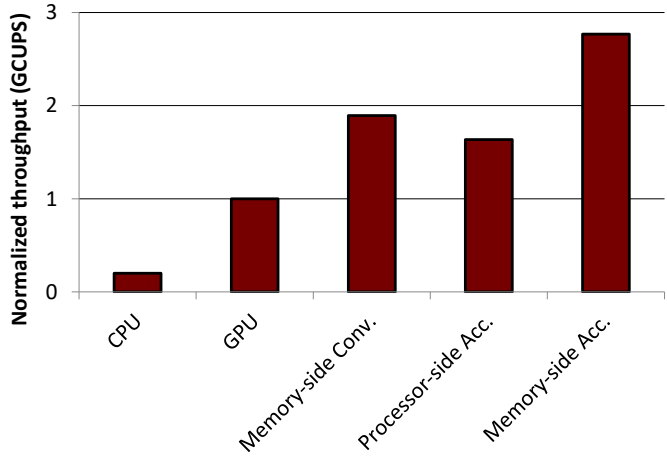

(a)

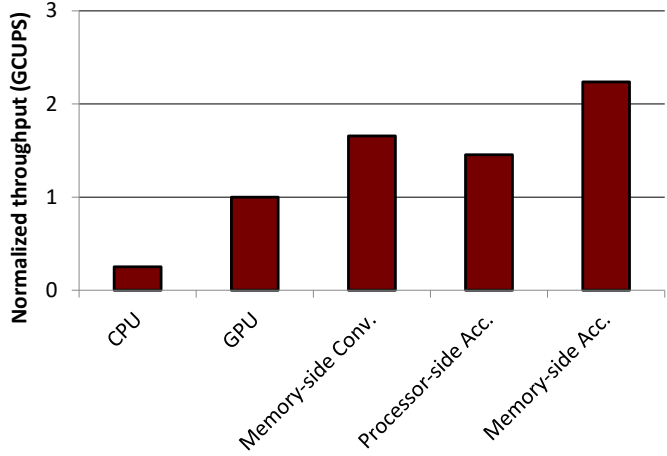

(b)

Figure 8. Throughput comparison under the (a) biological database search with Needleman-Wunsch and (b) short read assembly workloads. The numbers are normalized to the GPU results. For reference, the throughput of GPU is 15.9 GCUPS for database search and 19.8 GCUPS for assembly.

A Customized Memory-aware Architecture for Biological Sequence Alignment

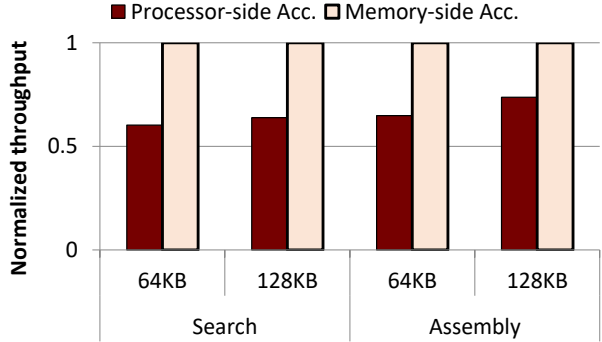

Figure 9. Throughput comparison for cache sizes

Figure 9 studies the impact of doubling the cache size of the accelerators on the obtained improvement.
As the figure indicates, the cache has a greater impact on the performance of the assembly workload. The reason is that the short read assembly workload works on shorter sequences, which can fit into reasonably-sized data caches. For the pairwise sequence alignment, as explored in Section 4, the large sequence size renders the cache less effective. This sublinear performance gain does not justify the area and power overheads of using more sophisticated multi-level cache structures.

b. *Sensitivity to the internal and external bandwidths*

The main source of speedup in our design is the much higher internal memory bandwidth at the memory-side compared to the external bandwidth available through the off-chip links.
The available internal and external bandwidths can vary significantly in different configurations. In particular, processors are required to provide four 16-bit high-speed interfaces in order to take advantage of the full external bandwidth of an HMC module (240 GB/s). This may lead to increasing the cost, area, and power of a processor.
To study the effect of the ration of the internal to external bandwidths on the benefits of our design, we fix the internal bandwidth to 320 GB/s and set the external bandwidth by 1x, 2x, and 4x lower.
The processor-side accelerators have 64KB data cache. As we expect, the speedup increases proportional to the bandwidth ratio and as Figure 10 demonstrates, reaches to 3.8x when the internal bandwidth is 4 times higher. Again, the cache can partially bridge the gap between the memory-side and processor-side designs for shorter sequences of the assembly process, but fall short in reducing the stress on the off-chip bandwidth for long sequences, on which the database search works.

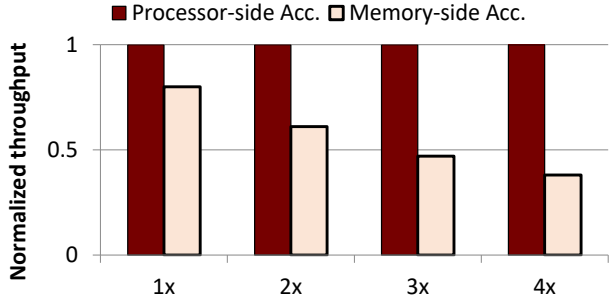

Figure 10. Throughput comparison for different internal to external bandwidth ratios

c. *Power evaluation*

Figure 11 shows the power consumption in the two memory-side and processor-side configurations under the database search application. The power consumption of the PEs are extracted from the hardware implementation of the designs (as described in Section 7) and the memory access power consumption for memory-side and processor-side PEs are set to 3.7 pJ/bit and 10 pJ/bit, respectively [14]. These energy numbers include the vault controller energy and the energy consumed to access the DRAM arrays for reads and writes. For external accesses, the packetization and link traversal energy are also included.

Figure 11 shows that the proposed PIM-based implementation can also outperform the processor-side design in terms of power consumption. The main source of power reduction in the proposed architecture is the elimination of on-board data transfer that has a considerable contribution to the total power consumption of computers [44].

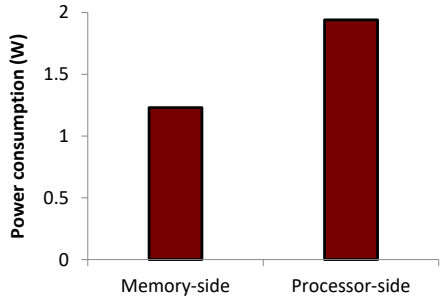

Figure 11. The power consumption comparison of PEs

## 9 CONCLUSION

In this paper, we proposed a memory-aware architecture to accelerate bioinformatics algorithms. Since biological sequence alignment is the base of the many algorithms in bioinformatics, this paper targeted accelerating the well-known Needleman–Wunsch sequence alignment. The main motivation behind the proposed architecture is the low operation density of sequence alignment algorithms that makes them very sensitive to memory bandwidth. The proposed architecture could manage to reduce the memory bandwidth demand by reusing the data and forwarding the intermediate results in a stream-based manner. To further push the bandwidth wall to increase throughput, the proposed architecture was implemented as a processing-in-memory unit at the logic

A Customized Memory-aware Architecture for Biological Sequence Alignment

layer of a 3D DRAM chip. The experimental results showed that moving the processing elements to the memory side leads to more than 60% improvement in throughput and 37% reduction in power consumption for the pairwise sequence alignment. Other bioinformatics algorithms that use alignment as part of their calculations can also benefit from the proposed architecture. As an example, we showed that the genome assembly problem can achieve to more 53% throughput on the proposed architecture. Compared to a GPU-based and a CPU-based design, the proposed architecture achieved 2.4X and 11.3X higher average throughput, respectively.